\documentclass[preprint,12pt]{elsarticle}
\usepackage{fullpage}
\usepackage{latexsym}
\usepackage[T1]{fontenc}
\usepackage{times}
\usepackage{amsmath}
\usepackage{amssymb}
\usepackage{amscd}
\usepackage{amsfonts}
\usepackage{amstext}
\usepackage{amsthm}
\usepackage{graphicx}
\usepackage[all]{xy}
\usepackage{float}
\usepackage{multirow}
\usepackage{color}
\usepackage{subcaption}

\begin{document}

\begin{frontmatter}

\title{On the phase diagram of the anisotropic XY chain in transverse magnetic field}
\author[tm]{Tomasz Maci\k{a}\.{z}ek }
\ead{maciazek@cft.edu.pl}
\author[jw]{Jacek Wojtkiewicz\corref{cor}}
\ead{wjacek@fuw.edu.pl}

\cortext[cor]{Corresponding author}
\address[tm]{Center for Theoretical Physics, Polish Academy of Sciences, Al. Lotnik\'ow 32/46, 02-668 Warszawa, Poland}
\address[jw]{Department for Mathematical Methods in Physics, Faculty of Physics, Warsaw University, ul. Pasteura 5, 02-093  Warszawa, Poland}

\begin{abstract}
We investigate an explicite formula for ground state energy of the anisotropic XY chain in 
transverse magnetic field.  In particular, we examine the smoothness properties
of this expression.  We  explicitly demonstrate that
the ground-state energy is infinitely differentiable on the boundary
 between ferromagnetic and oscillatory phases. We also confirm known 2d-Ising type behaviour
in the neighbourhood of certain lines of phase diagram and give more detailed information there, calculating a
 few next-to-leading exponents as well as the corresponding amplitudes.

\end{abstract}

\end{frontmatter}

\section{Introduction}
The quantum XY spin chain and its extensions have been studied for a very long time and from many different perspectives. It is due to a couple of reasons. First of all, it is possible
to obtain an exact solution (for spin one-half case) in the language of non-interacting
fermions \cite{LSM}. It is interesting for its own, and moreover 
it can be used in testing various techniques that are
applied to a wide range of non-integrable systems \cite{Henkel}.
Another motivation is to use the model to describe the experimental data of 
quasi-one-dimensional systems \cite{BunderMcKenzie}.
And finally, in recent years the XY chain has been extensively examined in the context of quantum information theory, quantum entropy and entanglement \cite{entropiaR} --  \cite{qInf5}.

The solution of the quantum spin one-half XY model is based on the Jordan-Wigner
transformation. By means of this transformation, the Hamiltonian can be brought to a form that describes a system of non-interacting fermions. This method has been used
in seminal paper by Lieb, Schulz and Mattis \cite{LSM}, where the anisotropic XY model without magnetic
field was studied. Another important papers generalizing and extending the results
of \cite{LSM} are: an exact solution of quantum Ising chain in transverse magnetic field
\cite{Pfeuty}, the ground-state phase diagram of the anisotropic XY model in transverse
magnetic field \cite{BCD, BCii},  the explicit expression for the ground-state energies \cite{AfterStolze},
 extensive computations of the correlation functions \cite{VT, JMea}, a simple and short derivation of the formulae for numerous observables \cite{Karevski}, the
exact results for XYZ model in magnetic field \cite{KurmannEtAl}.

\begin{figure}[H]
\centering
\includegraphics[width=.7\textwidth]{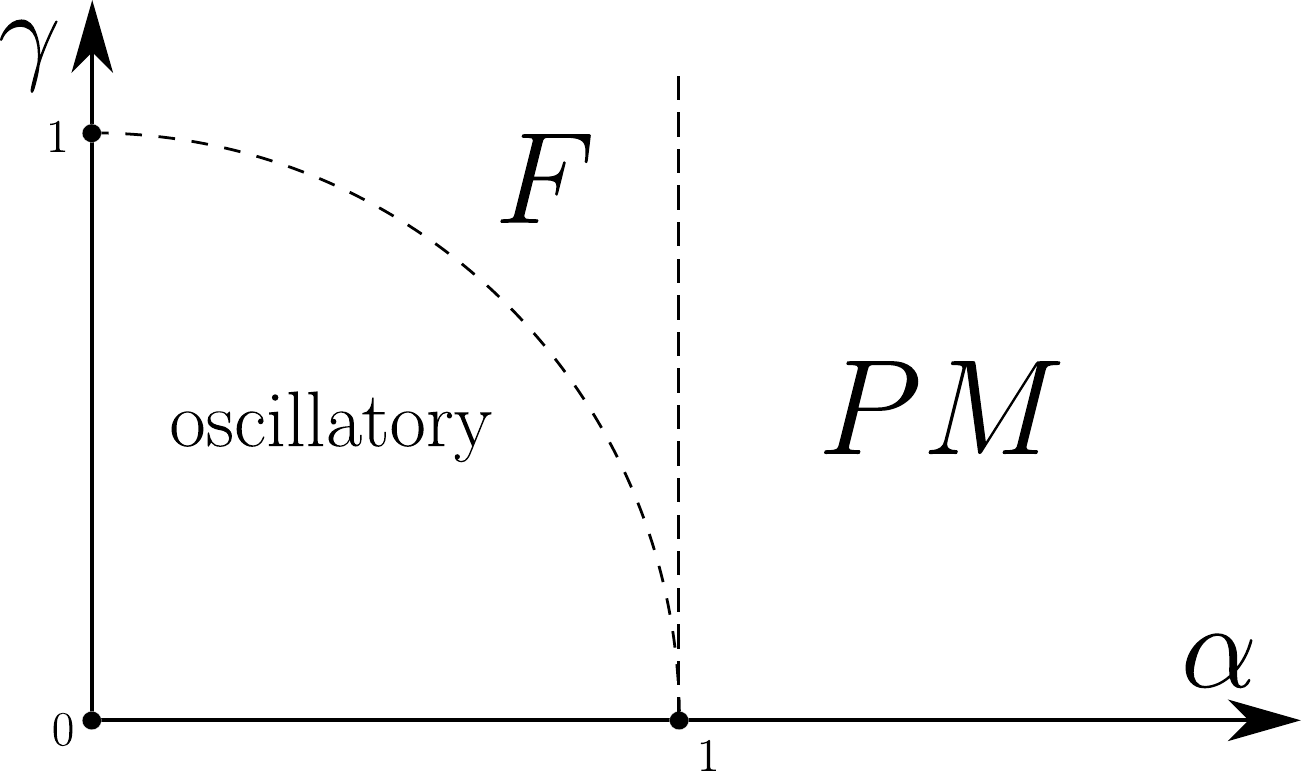}
\caption{The phase diagram of the considered ferromagnetic XY chain. According to the character of the two-spin correlations,
 one can distinguish three phases, i.e. a ferromagnetic phase ($F$),
 which splits into oscillatory and non oscillatory regions, and a paramagnetic phase ($PM$).
 Parameters $\alpha$ and $\gamma$ are the anisotropy parameter and the magnetic field strength from equation (\ref{ham1}).}
\label{PhaseDgm}
\end{figure}

The basic aim of our paper is to examine the {\em differentiability properties} of the expression for the ground state energy, which boils down to the study of the ground-state phase transitions' order.
This formula, expressible in terms of elliptic integrals
 has been known for quite a long time \cite{AfterStolze}, but some physical corollaries are, to our best knowledge, lacking.
Let us next present a detailed study of these aspects of the ground state formula in comparison to the phase diagram of
the XY model \cite{BCii}. The phase diagram is shown on Fig. \ref{PhaseDgm}. A good reference regarding the origin and nature of all phases (especially the oscillatory phase)  is \cite{KurmannEtAl, HvGR}.
 The most interesting lines
on the phase diagram are: {\em i)} the vertical line $\alpha=1$,
{\em ii)} the 'isotropy' line $\gamma=0$ and {\em iii)} the circle $\alpha^2+\gamma^2=1$.
(Here $\alpha$ is magnetic field, and $\gamma$ -- anisotropy parameter; 
proper definitions are given in Sec.~\ref{sec:HamAndDiag}.)
The first line {\em i)} has been exhaustively examined; it corresponds to the transition between the
paramagnetic and ferromagnetic phases.
 This transition is in the universality class of
two-dimensional classical Ising model \cite{Henkel, BunderMcKenzie}.
The second line {\em ii)} corresponds to the change of the direction of magnetization 
and is in the universality class of two decoupled Ising models \cite{BunderMcKenzie}.

The third line corresponds to the transition between the
ferromagnetic and oscillatory phases.
In the literature, some rather vague assertions concerning the nature of this 
transition have been formulated:  'the
thermodynamic functions do not exhibit singularities on this line' \cite{denNijs}, but no derivation
has been given. The expression for the ground state energy allows one to deduce that it is {\em infinitely differentiable} on this line. In this aspect, the transition between ferromagnetic and oscillatory phases
 resembles the Kosterlitz-Thouless one, but other aspects of these two transitions turned out to be 
 quite different.

The paper is organized as follows. In the Sec.~\ref{sec:HamAndDiag}
 we define the Hamiltonian and briefly describe 
its diagonalization, which leads to
an explicit expression for the ground-state energy. Its properties (differentiability and
asymptotic forms of solution in the neighbourhoods of phase transition lines)
are presented in the Sec. 3. The Sec. 4 contains summary of results obtained,
 comparison with existing
results and perspectives for future work.

%%%%%%%%%%%%%%%%%%%%%%%%%%%%%%%%%%%%%%%%%%%%%%%%%
\section{The Hamiltonian and its diagonalization}
\label{sec:HamAndDiag}
%%%%%%%%%%%%%%%%%%%%%%%%%%%%%%%%%%%%%%%%%%%%%%%%%

We consider a spin-$1/2$ chain consisting of $N$ sites. The spin sitting in a given site interacts
 only with its nearest neighbours. The Hamiltonian of such an interaction is given by $H_{i,j}=J_x S^x_i S^x_{j}+J_yS^y_i S^y_{j}$,
 where $i,j$ are the nearest neighbours; we denote it by 
 writing $\langle i,j\rangle$.  Moreover, the chain is placed in a transverse magnetic field $h$. Therefore, the Hamiltonian of the whole system is 
\begin{equation}
H=\sum_{\langle i,j\rangle} (J_x S_i^x S_{j}^x+J_y S_i^y S_{j}^y)-h\mu \sum_{i=1}^N S_i^z.
\label{ham0}
\end{equation}
The spin operators, $S^x$, $S^y$ and $S^z$ are proportional to the Pauli matrices
\begin{equation}
S^x=\frac{1}{2}
\begin{pmatrix}
0 & 1 \\
1 & 0 \\
\end{pmatrix},
\ 
\ S^y=\frac{1}{2}
\begin{pmatrix}
0 & -i \\
i & 0 \\
\end{pmatrix},
\ 
\ S^z=\frac{1}{2}
\begin{pmatrix}
1 & 0 \\
0 & -1 \\
\end{pmatrix}.
\end{equation}
We rewrite expression (\ref{ham0}) introducing the anisotropy parameter $\gamma\in[-1,1]$, defined by: $\gamma=\frac{1}{2J}(J_x-J_y)$,
where $J=\frac{1}{2}(J_x+J_y)$.
%In order to describe the anisotropy of the interaction between spins, we introduce parameter $\gamma\in[0,1]$. 
%and $J\in \mathbb{R}$.
 Moreover, we denote $h\mu$ as $\alpha$. The Hamiltonian (\ref{ham0}), expressed in terms of those parameters is given by
\begin{equation}
H=J\sum_{\langle i,j\rangle} \left[ (1+\gamma) S_i^x S_{j}^x+(1-\gamma) S_i^y S_{j}^y\right]-\alpha \sum_{i=1}^N S_i^z.
\label{ham1}
\end{equation}
For $\gamma=0$ Hamiltonian (\ref{ham1}) describes fully isotropic interaction, whereas $\gamma=1$ corresponds to the Ising model.
% The sign of the real parameter $J$ determines whether the interaction is ferromagnetic ($J<0$) or antiferromagnetic ($J>0$).
% From now on, we will restrict ourselves to considering $J=-1$, as the generalization to an arbitrary $J$ is straightforward.
We consider the ferromagnetic case, i.e. $J_x<0$, $J_y<0$.

%%%%%%%%
%%%%%%%%
%%%%%%%

%The Hamiltonian (\ref{ham1}) is 
%The main goal of this Section is to find the rigorous expression for the ground-state energy per spin
%of a system described by Hamiltonian (\ref{ham1})
% in the thermodynamic limit ($N\rightarrow \infty$). To this end, we apply the The Hamiltonian (\ref{ham1}) is
 %\cite{LSM}, \cite{Mattis}  and calculate the resulting integral. 

The Hamiltonian (\ref{ham1}) can be rewritten in the language of free fermions. It is done by the so called 
{\em Jordan-Wigner} transformation \cite{LSM}, \cite{Mattis}. For the free-fermion Hamiltonian one can explicitely
calculate all eigenvalues and in particular obtain the expression for the ground-state energy.

%%%%%%%%%%%
\subsection{Jordan-Wigner transformation and diagonalization of the Hamiltonian}
%%%%%%%%%%%

\noindent In thus section we give the main points of the Jordan-Wigner transformation. The details can be found for instance 
in \cite{LSM}, \cite{Mattis}. 
The first step is to write hamiltonian (\ref{ham1}) with $J=-1$ in terms of spin-raising and spin-lowering operators
\begin{equation*}
 a_i^\dagger=S_i^x+iS_i^y,\ \ a_i=S_i^x-iS_i^y.
\end{equation*}
The hamiltonian then reads
\begin{equation}
 H=-\sum_{(i,j)} \frac{1}{2}\left(a_i^\dagger a_{j}+\gamma a_i^\dagger a_{j}^\dagger+ \textrm{h.c.}\right)+\alpha\sum_{j=1}^N \left(a_i^\dagger a_i-\frac{1}{2}\right)
\label{ham_a}
\end{equation}
The algebraic properties of operators $(a_i)$ are nonuniform, i.e. 
the raising and lowering operators from the same site anticommute, 
while operators from different sites commute. The solution of this problem is the Jordan-Wigner transformation, which introduces the set of $2N$ fermionic operators $\{c_i\}$,
$\{c^\dagger_i\}$ given by
\begin{equation}
 \begin{cases}
  c_i=\exp\left(\pi i \sum_{j=1}^{i-1} a_j^\dagger a_j\right)a_i \\
  c_i^\dagger=a_i^\dagger \exp\left(-\pi i \sum_{j=1}^{i-1} a_j^\dagger a_j\right) \\ 
 \end{cases}
\end{equation}
The operators $c_i$ and $c_i^\dagger$ satisfy purely fermionic commutation rules, i.e. 
\begin{equation}
\{ c_i,c_j^{\dagger}\}=\delta_{ij},\,\,\{ c_i^\dagger,c_j^\dagger\}=\{ c_i,c_j\}=0.
\end{equation}
The Hamiltonian of the open chain then reads
\begin{gather}
 H=-\frac{1}{2}N\alpha-\sum_{i=1}^N \left(\frac{1}{2}\left[c_i^\dagger c_{i+1}+\gamma c_i^\dagger c_{i+1}^\dagger+\textrm{h.c}\right]-\alpha c_i^\dagger c_i\right)
%+ \nonumber \\ +\frac{1}{2}\left(c_N^\dagger c_{1}+\gamma c_N^\dagger c_{1}^\dagger+h.c.\right)
%\left(\exp\left(i\pi \mathfrak{N}\right)+1\right).
\label{ham_open}
\end{gather}
%One obtains the Hamiltonian in the case of the open chain simply by ommiting the terms with $(i,j)=(N,1)$ in (\ref{ham_cyclic}).

In calculating the ground state energy in the limit of large $N$, one can impose boundary conditions that are more convenient from the computational
point of view,
%neglect the boundary term proportional to $\exp\left(i\pi \mathfrak{N}\right)+1$ \cite{LSM}, \cite{Mattis}
 and therefore obtain the {\it c-cyclic chain} \cite{LSM}, \cite{Mattis}
\begin{equation}
 H=-\frac{1}{2}N\alpha-\sum_{i=1}^N \left(\frac{1}{2}\left[c_i^\dagger c_{i+1}+\gamma c_i^\dagger c_{i+1}^\dagger+\textrm{h.c}\right]-\alpha c_i^\dagger c_i\right),\,\,N+1\equiv1.
\label{ham_c}
\end{equation}
The hamiltonians (\ref{ham_open}), (\ref{ham_c}) are quadratic forms in fermion creation and annihilation operators and
 can be diagonalised, i.e. written in the following form
\begin{equation}
H=E_g+\sum_{k=1}^{N}\Lambda_k\eta^\dagger_k\eta_k,
\label{ham_diag}
\end{equation}
where $(\eta_k)$ are some fermionic operators.
Assume, we have found the eigenvalues $\Lambda_k$ of Hamiltonian (\ref{ham_a}).
 In order to find the ground state energy, $E_g$, let us calculate the trace of Hamiltonians 
 (\ref{ham_a}) and (\ref{ham_diag})
\begin{equation}
 \begin{array}{c}
  \mathrm{Tr}(H)=2^{N-1}\sum_k \Lambda_k+2^N E_g \\
\hspace{2.9cm}=\alpha\mathrm{Tr}\left\{\sum_{j=1}^N \left(a_i^\dagger a_i-\frac{1}{2}\right)\right\}=0.
 \end{array}
\label{trace}
\end{equation}
 \noindent The formula for the ground state energy per spin in the thermodynamic limit does not depend on the boundary conditions,
  because the difference between the c-cyclic and open chain 
 % is non-zero only for finite number of terms and 
tend to zero in the thermodynamic limit.

\noindent Therefore, in all cases, the ground state energy can be calculated from equation (\ref{trace}), i.e.
\begin{equation}
E_g=-\frac{1}{2}\sum_k \Lambda_k.
\label{eg_sum}
\end{equation}
The eigenvalues for the cyclic boundary conditions are given by 
\begin{equation}
\Lambda_k^2=\left[\alpha-\cos(k)\right]^2+\gamma^2\sin^2(k),\ k=\frac{2\pi m}{N},\ m=0,1,\dots,N-1.
\label{Eigenvs}
\end{equation}
This expression was derived for the first time by Katsura in \cite{Katsura}.

%%%%%%%%%%%
\subsection{The formulae for ground state energy}
%%%%%%%%%%%

\begin{figure}[H]
\centering
\includegraphics[width=.7\textwidth]{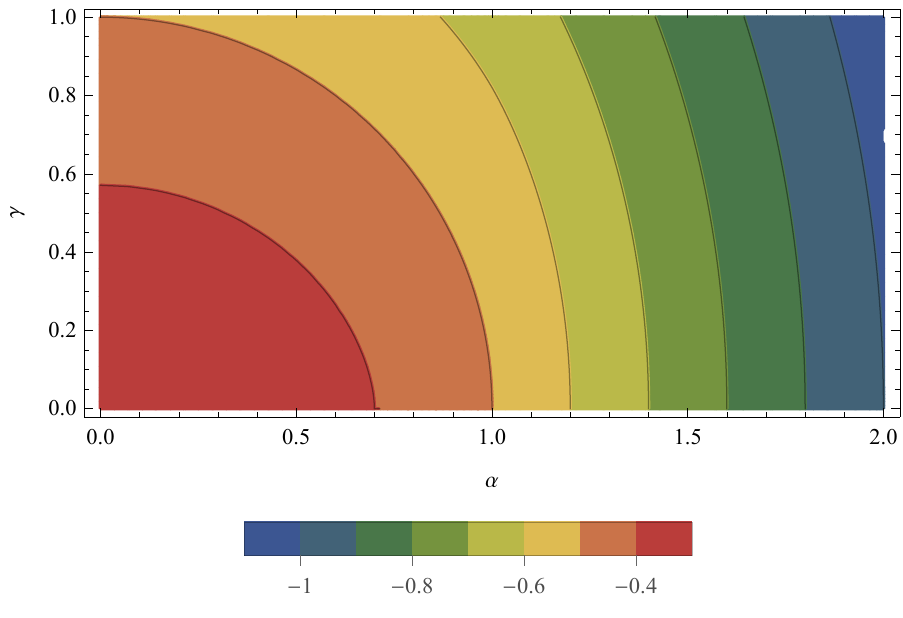}
\caption{The ground state energy map. The ground state energy is a continuous function with one maximum at $(\alpha,\gamma)=(0,0)$. One characteristic circle-shaped level set can be seen. This is the circle $\alpha^2+\gamma^2=1$, where $E_g=-\frac{1}{2}$.}
\label{energy_map}
\end{figure}

Using fomulae (\ref{eg_sum}) and (\ref{Eigenvs}) in the thermodynamic limit, one obtains the ground state energy in a form of the following integral
\begin{equation}
\frac{E_g}{N}=-\frac{1}{4\pi} \int_0^{2\pi}dt\sqrt{(\alpha-\cos t)^2+\gamma^2 \sin^2 t}=-\frac{1}{2\pi} \int_0^{\pi}dt\sqrt{(\alpha-\cos t)^2+\gamma^2 \sin^2 t}.
\label{gs_int}
\end{equation}
The procedure of calculation of this integral is lengthy and tedious but straightforward. We skip it and present the final result \cite{AfterStolze}
\footnote{We have derived an expression for $\epsilon_g$ being not aware of paper \cite{AfterStolze}. After the first version of our manuscript was completed, 
Prof. J. Stolze drew our attention to this paper.}:
The ground state energy $\epsilon_g$ is given by formulas
\begin{footnotesize}
\begin{equation}
\epsilon_g=
\left\{
\begin{array}{cc}
-\frac{\sqrt{1-\alpha^2}}{\pi}\Bigg{[}E(k)-K(k)+\frac{1}{1-\alpha^2}\Pi\bigg{(}-\frac{\alpha^2}{1-\alpha^2};k\bigg{)}\Bigg{]}\;,\;\;\;k=\sqrt{\frac{1-\alpha^2-\gamma^2}{1-\alpha^2}}, & \mathrm{for}\ \alpha^2+\gamma^2<1 \\
-\frac{1-\alpha^2}{\pi\gamma}\Bigg{[}\Pi\big{(}\alpha^2;k\big{)}-K(k)+\frac{\gamma^2}{1-\alpha^2}E(k)\Bigg{]}\;,\;\;\;k=\frac{\sqrt{\alpha^2+\gamma^2-1}}{\gamma}, & \mathrm{for}\ \alpha^2+\gamma^2>1,\ \alpha<1 \\
-\frac{\alpha^2-1}{\pi\sqrt{\alpha^2+\gamma^2-1}}\Bigg{[}\Pi\big{(}\frac{1}{\alpha^2};k\big{)}-K(k)+\frac{\alpha^2+\gamma^2-1}{\alpha^2-1}E(k)\Bigg{]}\;,\;\;\ k=\frac{\gamma}{\sqrt{\alpha^2+\gamma^2-1}} & \mathrm{for}\ \alpha^2+\gamma^2>1,\ \alpha>1.
\end{array}
\right.
\label{eg_final}
\end{equation}
\end{footnotesize}
\noindent The above results are formulated in terms of complete elliptic integrals \cite{Fichtenholz}:
\begin{gather}
K(k)=\int_0^1\frac{dz}{\sqrt{(1-z^2)(1-k^2z^2)}}, \nonumber \\
E(k)=\int_0^1dz\sqrt{\frac{1-k^2z^2}{1-z^2}}, \label{elliptic} \\ 
\Pi(n;k)=\int_0^1\frac{dz}{(1-nz^2)\sqrt{(1-z^2)(1-k^2z^2)}}. \nonumber
\end{gather}
which are called complete elliptic integrals of the first, second and the third kind, respectively. To make the expression more transparent, we plot the function (\ref{eg_final})  on the Fig. \ref{energy_map}.

%%%%%%%%%%%%%%%%%%%%%%%%%%%%%%%
\section{Thermodynamic functions near the phase transition lines}
%%%%%%%%%%%%%%%%%%%%%%%%%%%%%%%
\subsection{Magnetization and magnetic susceptibility}
\label{section:magn_susc}
\begin{figure}[H]
\centering
\begin{subfigure}{.47\textwidth}
  \centering
  \includegraphics[width=\textwidth]{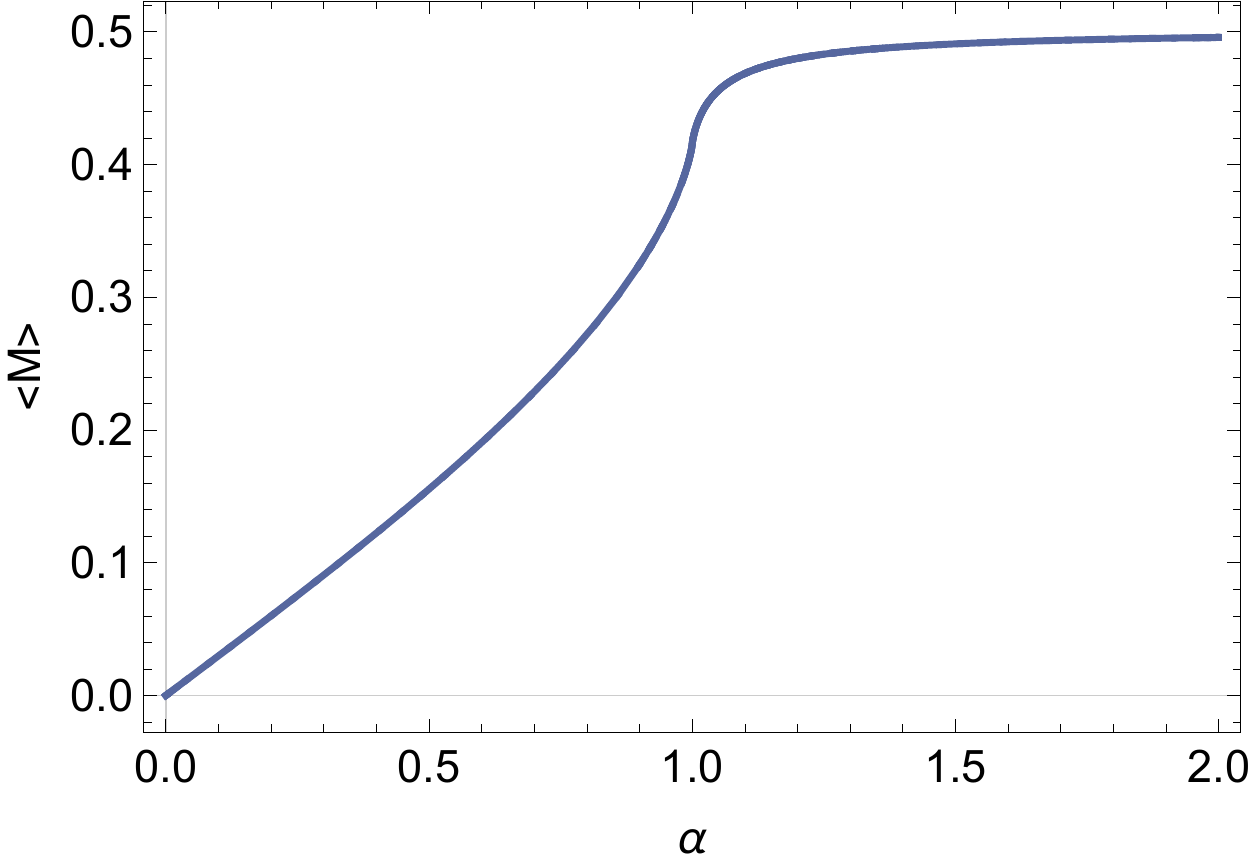}
  \caption{}
  \label{magn}
\end{subfigure}
\hspace{5mm}
\begin{subfigure}{.47\textwidth}
  \centering
  \includegraphics[width=\textwidth]{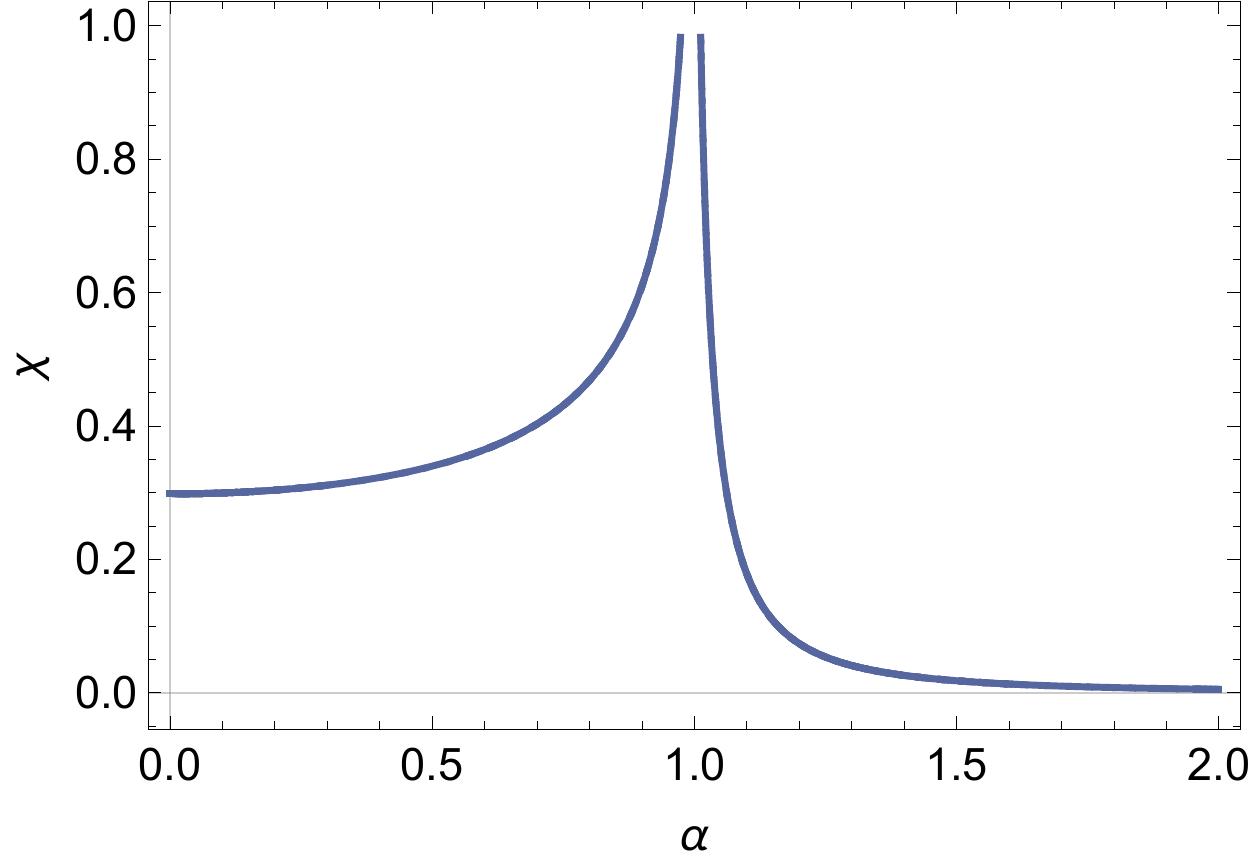}
  \caption{}
  \label{susc}
\end{subfigure}
\caption{Magnetisation (a) and magnetic susceptibility (b) for $\gamma=1/3$.}
\label{magn_and_susc}
\end{figure}

The free energy per particle of the system described by a Hamiltonian $H$ is of the form $f=-\frac{1}{N\beta}\ln Z$,
 where $Z=\mathrm{Tr}e^{-\beta H}$ and $\beta$ is the inverse
temperature. When the system is in its ground state, one has to calculate the limit of $f$ as $\beta$ approaches infinity. Therefore
\begin{equation*}
f_g=\lim_{\beta\rightarrow\infty}\frac{1}{N\beta}\ln Z=\frac{1}{N}\lim_{\beta\rightarrow\infty}\frac{1}{\beta}\mathrm{Tr}e^{-\beta H}=\epsilon_g.
\end{equation*}
Then the (transverse) magnetization and the magnetic susceptibility are simply given by
\begin{gather}
\langle M \rangle=-\frac{\partial \epsilon_g}{\partial \alpha},\ \ \ \chi=-\frac{\partial^2 \epsilon_g}{\partial \alpha^2}.
\label{magn_susc}
\end{gather}
Differentiating equations (\ref{eg_final}), one obtains 
\begin{footnotesize}
%\small{
\begin{equation}
\langle M \rangle=
\left\{
\begin{array}{cc}
\frac{1}{\pi\alpha\sqrt{1-\alpha^2}}\Big{[}\Pi\Big{(}-\frac{\alpha^2}{1-\alpha^2};k\Big{)}-(1-\alpha^2)K(k)\Big{]}\;,\;\;\;k=\sqrt{\frac{1-\alpha^2-\gamma^2}{1-\alpha^2}}, & \mathrm{for}\ \alpha^2+\gamma^2<1 \\
\frac{1-\alpha^2}{\pi\alpha\gamma}\Big{[}\Pi\big{(}\alpha^2;k\big{)}-K(k)\Big{]}\;,\;\;\; k=\frac{\sqrt{\alpha^2+\gamma^2-1}}{\gamma}, & \mathrm{for}\ \alpha^2+\gamma^2>1,\ \alpha<1 \\
\frac{\alpha^2-1}{\pi\alpha\sqrt{\alpha^2+\gamma^2-1}}\Pi\big{(}\frac{1}{\alpha^2};k\big{)}\;,\;\;\; k=\frac{\gamma}{\sqrt{\alpha^2+\gamma^2-1}} & \mathrm{for}\ \alpha^2+\gamma^2>1,\ \alpha>1.
\end{array}
\right.
\label{magnetization}
\end{equation}
%}%end{small}
\end{footnotesize}
and
\begin{equation}
\chi=
\left\{
\begin{array}{cc}
\frac{(1-\alpha^2)E(k)-\gamma^2K(k)}{\pi\sqrt{1-\alpha^2}(1-\alpha^2-\gamma^2)}\;,\;\;\; k=\sqrt{\frac{1-\alpha^2-\gamma^2}{1-\alpha^2}}, & \mathrm{for}\ \alpha^2+\gamma^2<1 \\
\frac{\gamma[K(k)-E(k)]}{\pi(\alpha^2+\gamma^2-1)}\;,\;\;\; k=\frac{\sqrt{\alpha^2+\gamma^2-1}}{\gamma}, & \mathrm{for}\ \alpha^2+\gamma^2>1,\ \alpha<1 \\
-\frac{K(k)-E(k)}{\pi\sqrt{\alpha^2+\gamma^2-1}}\;,\;\;\; k=\frac{\gamma}{\sqrt{\alpha^2+\gamma^2-1}} & \mathrm{for}\ \alpha^2+\gamma^2>1,\ \alpha>1.
\end{array}
\right.
\label{susceptibility}
\end{equation}
The exemplary plots are shown on Figure. \ref{magn_and_susc} 
\iffalse
\begin{figure}[H]
\centering
\subfloat[Magnetization for $\gamma=1/3$.]{\includegraphics[width=6.6cm,height=5cm]{magn.pdf}\label{magn_plot}} \hspace{5mm}
 \subfloat[Magnetic susceptibility for $\gamma=1/3$.]{\includegraphics[width=6.6cm,height=5cm]{chi.pdf}\label{susc_plot}}
\end{figure}
\fi

As one can see in figure \ref{susc}, the magnetic susceptibility approaches infinity near $\alpha=1$. Using the series expansions of the complete 
elliptic integrals near $k=1$ \cite{WW}, one has that for $|1-\alpha|$ small
\[
\chi = \frac{1}{\pi\gamma}\left[\log{\frac{1}{\sqrt{|1-\alpha|}}}+\log\left({2\sqrt{2}\gamma}\right)-1\right]+\frac{8-9\log2-\gamma^2-6\log\gamma}{4\pi\gamma^3}|1-\alpha|+ \
\]
\begin{equation}
 +\frac{3}{4\pi\gamma^3}|1-\alpha|\log|1-\alpha|+ \mathcal{O}(|1-\alpha|^2).
\label{Rozwiniecie1}
\end{equation}
Another singularity arises when the second derivative of the energy with
 respect to $\alpha$ is evaluated around $\gamma=0$ (line {\em ii)}). One may expect such a behaviour,
  because for $\gamma>0$ the x-coupling constant, $J_x$, is greater than the y-coupling constant, $J_y$ (see eqs. (\ref{ham0}) and (\ref{ham1})), while for $\gamma<0$
we have $J_x<J_y$.
   This  means that while crossing the line $\gamma=0$, the direction of magnetization changes from $x$ to $y$ direction. It turns out that such a transition 
    occurs as an Ising--type singularity, i.e. as a logarithmic divergence of the second derivative. Using the expansions of the elliptic integrals around $1$, $\infty$ or $0$, depending on the value of $k$ for $\gamma$ close to $0$, one can see that this is the case for $\alpha<1$ and $|\gamma|$ small
\[
\frac{\partial^2 \epsilon_g}{\partial \gamma^2}=\frac{\sqrt{1-\alpha^2}}{\pi}\left[ 2-\frac{\alpha}{\sqrt{1-\alpha^2}}\arcsin\alpha+\log\left(\frac{|\gamma|}{4\sqrt{1-\alpha^2}}\right)\right]+\frac{1}{4\pi\sqrt{1-\alpha^2}}\Big(15-21\alpha^2+
\]
 \begin{equation}
 -18\alpha\sqrt{1-\alpha^2}\arcsin\alpha\Big)\gamma^2+\frac{9(1-2 \alpha^2)}{8\pi\sqrt{1-\alpha^2}}\log\left[\frac{\gamma^2}{16(1-\alpha^2)}\right]\gamma^2
+\mathcal{O}(|\gamma|^3).
\label{Rozwiniecie2}
\end{equation}
For $\alpha>1$ the ground state energy is smooth.

{\em Remark.} The transverse susceptibility diverges on both lines $\alpha=1$ and $\gamma=0$ in a logarithmic 
manner. But the behaviour of the correlation functions is different. The corresponding universality
classes for these transitions are called 'Ising model'
and 'two decoupled Ising models', respectively.  They are  governed by different
correlation functions' critical exponents \cite{BunderMcKenzie}.
These facts are well known. But using an exact solution, one can obtain the 
sub-leading exponents as well as amplitudes up to arbitrarily high order.
%We have presented above some of first such terms.
Equations (\ref{Rozwiniecie1}) and (\ref{Rozwiniecie2}) present the first terms of the expansion.

%%%%%%%%%%%%
\subsection{Smoothness of the ground state energy on line $\alpha^2+\gamma^2=1$}
In this section we will give the formula for all partial derivatives of function $\epsilon_g(\alpha,\gamma)$ (see equation (\ref{eg_final})) with respect to parameter $\alpha$ in the limit $\alpha\rightarrow\sqrt{1-\gamma^2}^\pm$. As shown in section \ref{section:magn_susc}, the second derivatives of $\epsilon_g$ are given by
\begin{equation}
\frac{\partial^2\epsilon_g}{\partial \alpha^2}=
\left\{
\begin{array}{cc}
\frac{(1-\alpha^2)E(k)-\gamma^2K(k)}{\pi\sqrt{1-\alpha^2}(1-\alpha^2-\gamma^2)}\; , \;\;\; k=\sqrt{\frac{1-\alpha^2-\gamma^2}{1-\alpha^2}}, & \mathrm{for}\ \alpha^2+\gamma^2<1 \\
-\frac{\gamma[E(k)-K(k)]}{\pi(\alpha^2+\gamma^2-1)}\; , \;\;\; k=\frac{\sqrt{\alpha^2+\gamma^2-1}}{\gamma}, & \mathrm{for}\ \alpha^2+\gamma^2>1,\ \alpha<1.
\label{2nd_long}
\end{array}
\right.
\end{equation}
We will next write the above formula in a more compact form, using identities
\begin{equation}
K(k)=\frac{1}{\sqrt{1-k^2}}K\left(\frac{ik}{\sqrt{1-k^2}}\right),\ \ E(k)=\sqrt{1-k^2}E\left(\frac{ik}{\sqrt{1-k^2}}\right)
\label{identities}
\end{equation}
which follow from equations 17.3.29 and 17.3.30 in \cite{AS}. For $k=\sqrt{\frac{1-\alpha^2-\gamma^2}{1-\alpha^2}}$, equations (\ref{identities}) read
\begin{equation*}
K(k)=\frac{\sqrt{1-\alpha^2}}{\gamma}K\left(\tilde k\right),\ \ E(k)=\frac{\gamma}{\sqrt{1-\alpha^2}}E\left(\tilde k\right),
\end{equation*}
where $\tilde k=\frac{ik}{\sqrt{1-k^2}}=i\frac{1}{\gamma}\sqrt{1-\alpha^2-\gamma^2}$. Inserting this result to formula (\ref{2nd_long}), one obtains that
\begin{equation*}
\frac{\partial^2\epsilon_g}{\partial \alpha^2}=-\frac{\gamma\left[E\left(\tilde k\right)-K\left(\tilde k\right)\right]}{\pi(\alpha^2+\gamma^2-1)}=-\frac{1}{\pi\gamma}\frac{\left[E\left(\tilde k\right)-K\left(\tilde k\right)\right]}{\tilde k^2},\ \  \alpha<1.
\end{equation*}
We will next use the identity
\begin{equation*}
2\frac{\partial E\left(k\right)}{\partial\left(k^2\right)}=\frac{\left[E\left(k\right)-K\left(k\right)\right]}{k^2},
\end{equation*}
which can be derived using definitions of complete elliptic integrals (equations (\ref{elliptic})) simply by subtraction of the integrands on the right hand side and by differentiation under the integral sign on the left hand side. The second derivative of the ground state energy now reads
\begin{equation}
\frac{\partial^2\epsilon_g}{\partial \alpha^2}=-\frac{2}{\pi\gamma}\frac{\partial E\left(k\right)}{\partial\left(k^2\right)}\Bigg{|}_{k=\tilde k}.
\label{gse_2nd_der}
\end{equation}
Higher derivatives of the ground state energy can be therefore calculated as derivatives of the composite function $E\left(\tilde k^2(\alpha,\gamma)\right)$. To this end, we will use the following formula for the $n$th derivative of the composition of two functions \cite{HMY}
\begin{equation}
\frac{d^n}{dx^n}f(g(x))=\sum\frac{n!}{k_1!k_2!\dots k_n!}f^{(k)}(g(x))\left(\frac{g'(x)}{1!}\right)^{k_1}\left(\frac{g''(x)}{2!}\right)^{k_n}\dots \left(\frac{g^{(n)}(x)}{n!}\right)^{k_n},
\label{nthderivative_gen}
\end{equation}
where the sum runs through all positive integers $k_1,\dots,k_n$ such that $k_1+2k_2+\dots+nk_n=n$ and $k:=k_1+k_2+\dots+k_n$. Note that in the case at hand, the third derivarive of $g(x)=\frac{\partial^3\left(\tilde k^2\right)}{\partial\alpha^3}$ vanishes, hence equation (\ref{nthderivative_gen}) simplifies. Namely,
\begin{equation}
\frac{d^n}{dx^n}f(g(x))=\sum_{k=0}^{\lfloor \frac{n}{2}\rfloor} \frac{n!}{k!(n-2k)!}f^{(n-k)}\left(g(x)\right)\left(g'(x)\right)^{n-2k}\left(\frac{1}{2}g''(x)\right)^k.
\label{n_der}
\end{equation}
Combining the above formula with equation (\ref{gse_2nd_der}), one obtains
\begin{equation}
\lim_{\alpha\rightarrow\sqrt{1-\gamma^2}^\pm}\frac{\partial^{n+2}\epsilon_g}{\partial \alpha^{n+2}}=-\frac{2}{\pi\gamma}\sum_{k=0}^{\lfloor \frac{n}{2}\rfloor} \frac{2^{n-2k}n!}{k!(n-2k)!}\left(\frac{1}{\gamma^2}\right)^{n-k}\alpha^{n-2k}\frac{\partial^{n+2}E\left(k\right)}{\partial\left(k^2\right)^{n+2}}\Bigg{|}_{k=0},
\end{equation}
where the derivatives of function $E(k)$ are known from its series expansion \cite{AS}
\begin{equation}
\frac{\partial^{m} E\left(k\right)}{\partial\left(k^2\right)^{m}}\Bigg{|}_{k=0}=\frac{\pi}{2}\left[\frac{(2m)!}{2^{2m}(m!)^2}\right]^2\frac{1}{1-2m}
\end{equation}
Both limits of derivatives: left and right ones are equal up to an arbitrary order, so the ground-state energy $\epsilon_g$ is infinitely differentiable on the circle $\alpha^2+\gamma^2=1$.

We have also checked the smoothness of the energy numerically, by plotting the derivatives of the ground state energy of order up to five along some sample curves passing transversally through the circle $\alpha^2+\gamma^2=1$. The results confirmed our calculations within numerical accuracy.

\subsection{The gap energy near the line $\alpha^2+\gamma^2=1$}
Because the character of correlations changes while crossing the line $\alpha^2+\gamma^2=1$, one could expect that it should result in some kind of singularity. However, as we have shown in previous paragraph, the ground state energy is smooth on the aforementioned circle. Such a behaviour resembles the {\em  Kosterlitz-Thouless transition}, where certain thermodynamic quantities are smooth. So one can ask whether this 
transition is present in the considered XY spin chain. Let us shortly review the main properties of the K-T  transition, necessary to verify this hypothesis. 

The Kosterlitz-Thouless transition occurs in many two-dimensional classical models: Coulomb gas, sine-Gordon, XY and many others \cite{KT}.
There is no spontaneous symmetry breaking and the free energy, although non-analytic, is infinitely differentiable during this transition.
The singular part of the free energy behaves as:
\[
f_{sing}\sim \exp\left(-\frac{C}{\sqrt{|T-T_c|}}\right)
\] 
near the critical temperature $T_c$ ($C$ is a positive constant). 
 The Kosterlitz-Thouless transition occurs also in ground states of one-dimensional quantum chains under change of certain parameter $\lambda$ 
(for instance, an anisotropy
parameter in the quantum XXZ chain \cite{Samaj}; the $O(2)$ quantum chain \cite{AltonHamer}; the Ashkin-Teller chain
 \cite{qATchain1}, \cite{qATchain2}; spin-1 Heisenberg models \cite{NeirottiOliveira}). In all cases,
 the ground-state energy is infinitely differentiable, and the energy gap $\Delta E$
over the 
ground state behaves near the critical value of parameter $\lambda_c$ as
\begin{equation}
\Delta E\sim \exp\left(-\frac{C}{\sqrt{|\lambda-\lambda_c|}}\right)
\label{KTinQChains}
\end{equation}
($C$ is a positive constant).
 The physical origin of the Kosterlitz-Thouless sometimes
is clear  \cite{AltonHamer}, but sometimes is somewhat hidden \cite{Samaj}, \cite{qATchain2}. So we decided to check whether the K-T transition
is present in the XY model.
In the following paragraph, we will investigate the aforementioned gap energy i.e. the difference between the energy of the first excited state and the energy of the ground state. First, we will reproduce in an alternative way the well known result regarding the periodic chain. Then, we will show the results of the numerical calculations for an open chain, which reveal that the behaviour of the energy gap is qualitatively the same in both cases. While the energy gap for the cyclic chain is well-studied, the open chain, to our best knowledge, has not yet been investigated. The main reason for this is the lack of the translational symmetry, which makes the exact diagonalization far more difficult. From equation (\ref{ham_diag}), one can easily see that the gap energy is given by the minimal eigenvalue of the Hamiltonian
\begin{equation}
\Delta_N(\alpha,\gamma)=\frac{1}{N}\min_{k}\Lambda_k.
\label{gap_general}
\end{equation}
Let us first reexamine the case of the $c$-cyclic boundary conditions. In order to find the minimal eigenvalue (equation (\ref{gap_general})) in the limit of $N\rightarrow\infty$, one has to solve the equation $\frac{d}{dk}\Lambda_k=0$, i.e.
\begin{equation*}
2\sin k\left(\alpha-\cos k+\gamma^2\cos k\right)=0.
\end{equation*}
The gap energy in then given by
\begin{equation*}
\lim_{N\rightarrow\infty}N\Delta_N(\alpha,\gamma)=
\begin{cases}
\frac{\gamma}{\sqrt{1-\gamma^2}}\sqrt{1-\alpha^2-\gamma^2}\  \\
|\alpha-1|\ 
\end{cases}
\begin{array}{c}
\textrm{for }\alpha\leq\sqrt{1-\gamma^2}, \\
\textrm{for }\alpha>\sqrt{1-\gamma^2}
\end{array}
\end{equation*}
and it is clear that it does not describe the Kosterlitz-Thouless transition. This result is 
consistent with the first-order of the expansion for the energy gap, which was derived in \cite{Henkel2}.

 Let us next switch our attention to the open boundary conditions. The hamiltonian of an open chain after the
 Jordan-Wigner transformation is given by (\ref{ham_open}). Because of the lack of the translational symmetry, the exact diagonalization of the fermionic Hamiltonian is more difficult, 
i.e. eigenvalues of one-particle problem are not analytically known for arbitrary $N$.
 However, one can do the numerical diagonalization. The numerical calculations show that also in the case of an open chain,
 the gap energy behaves like $1/N$ for sufficiently long chains. 
 The exemplary plots depicting this phenomenon are shown in Figure \ref{fig:gap_n}, where the line
\begin{equation}
\Delta_N=\frac{a}{N}+\Delta_\infty
\label{gap_fit}
\end{equation}
was fitted.

\begin{figure}[H]
\centering
\begin{subfigure}{.47\textwidth}
  \centering
  \includegraphics[width=\textwidth]{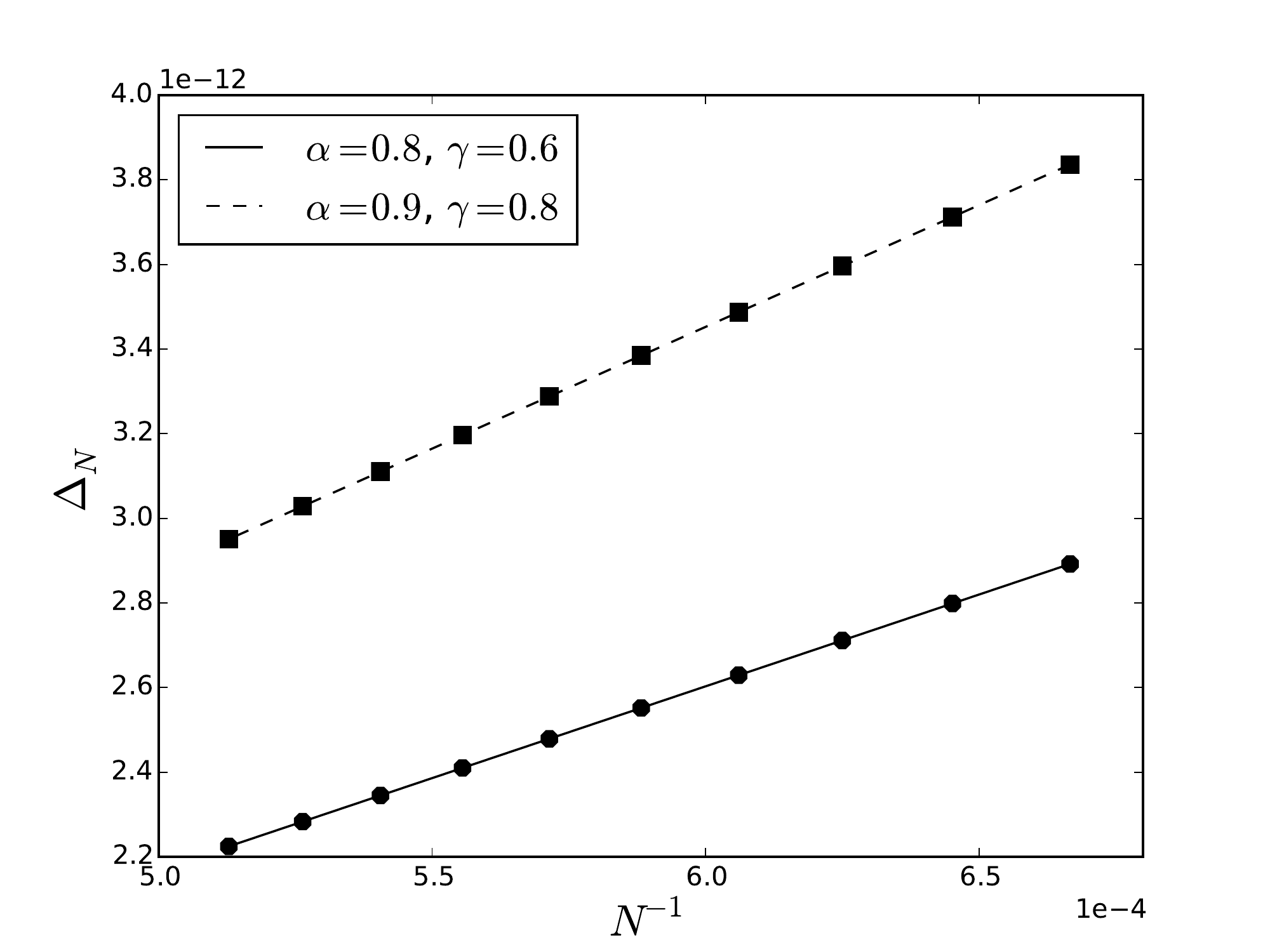}
  \caption{Fitted parameters are: $a=4.33\cdotp10^{-9},\ \Delta_\infty=(-1\pm2)\cdotp10^{-27}$ for $\alpha=0.8,\ \gamma=0.6$ and $a=5.75\cdotp10^{-9},\ \Delta_\infty=(-2\pm3)\cdotp10^{-27}$ for $\alpha=0.9,\ \gamma=0.8$.}
\end{subfigure}
\hspace{5mm}
\begin{subfigure}{.47\textwidth}
  \centering
  \includegraphics[width=\textwidth]{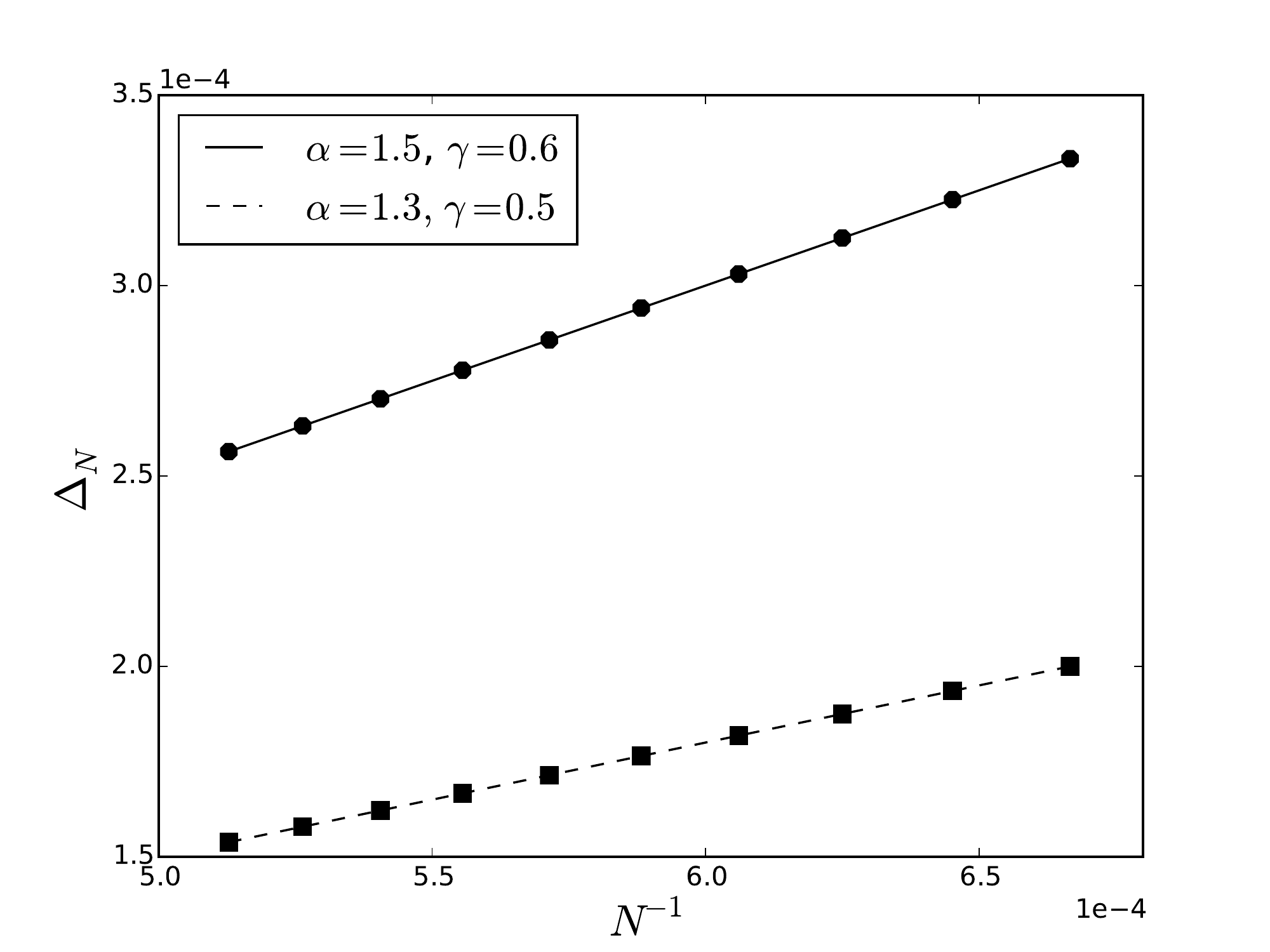}
  \caption{Fitted parameters are: $a=0.5,\ \Delta_\infty=(3.43\pm1.5)10^{-9}$ for $\alpha=1.5,\ \gamma=0.6$ and $a=0.3,\ \Delta_\infty=(3.65\pm1.6)10^{-9}$ for $\alpha=1.3,\ \gamma=0.5$.}
\end{subfigure}
\caption{The energy gap versus the chain length in (a) weak and (b) strong field.} 
\label{fig:gap_n}
\end{figure}

The fitted values of $\Delta_\infty$ are equal to zero within the accuracy of $10^{-27}$ and $10^{-9}$ respectively. The values of parameter $a$ are also fitted very accurately - up to the magnitude of $10^{-24}$ and $10^{-7}$ respectively.\par
Since $\Delta_\infty$ is zero in the whole range of parameters, there is no Kosterlitz-Thouless transition. However, the behaviour of $N\Delta_N$ in the limit or large $N$, i.e. parameter $a$ from equation (\ref{gap_fit}), varies significantly on the choice of the boundary conditions. However, in the strong-field region, $\alpha>1$, plots on figure (\ref{fig:gap_n}) suggest that $a=|\alpha-1|$ regardless on the choice of the boundary conditions.

It turns out that the behaviour of energy gap in the considered XY chain is {\em not} compatible with behaviour described by (\ref{KTinQChains}). It is so due to the following reasons.
\begin{itemize}
\item Consider the ordered phase, i.e.  $\alpha>\sqrt{1-\gamma^2}$ (and $\alpha<1$). In this ordered phase, the system exhibits the $\mathbb{Z}_2$ symmetry \cite{HvGR} (corresponding to plus and minus sign of magnetization), which implies the (asymptotic) degeneracy of two 
ground states. So, the energy gap is zero in the ordered phase (in the thermodynamic limit).
\item Consider the oscillating phase (i.e. $\alpha<\sqrt{1-\gamma^2}$) and take the line $\gamma=$const. It turns
out that along this line, the gap {\em oscillates}. Such a behaviour 
has been observed in  \cite{HvGR}; we illustrate it on Fig. \ref{fig:gap_n}. Moreover, for all values of $\alpha$, the gap
 tends to 0 in the thermodynamic limit. 
\end{itemize}

\begin{figure}[H]
\centering
\begin{subfigure}{.47\textwidth}
  \centering
  \includegraphics[width=\textwidth]{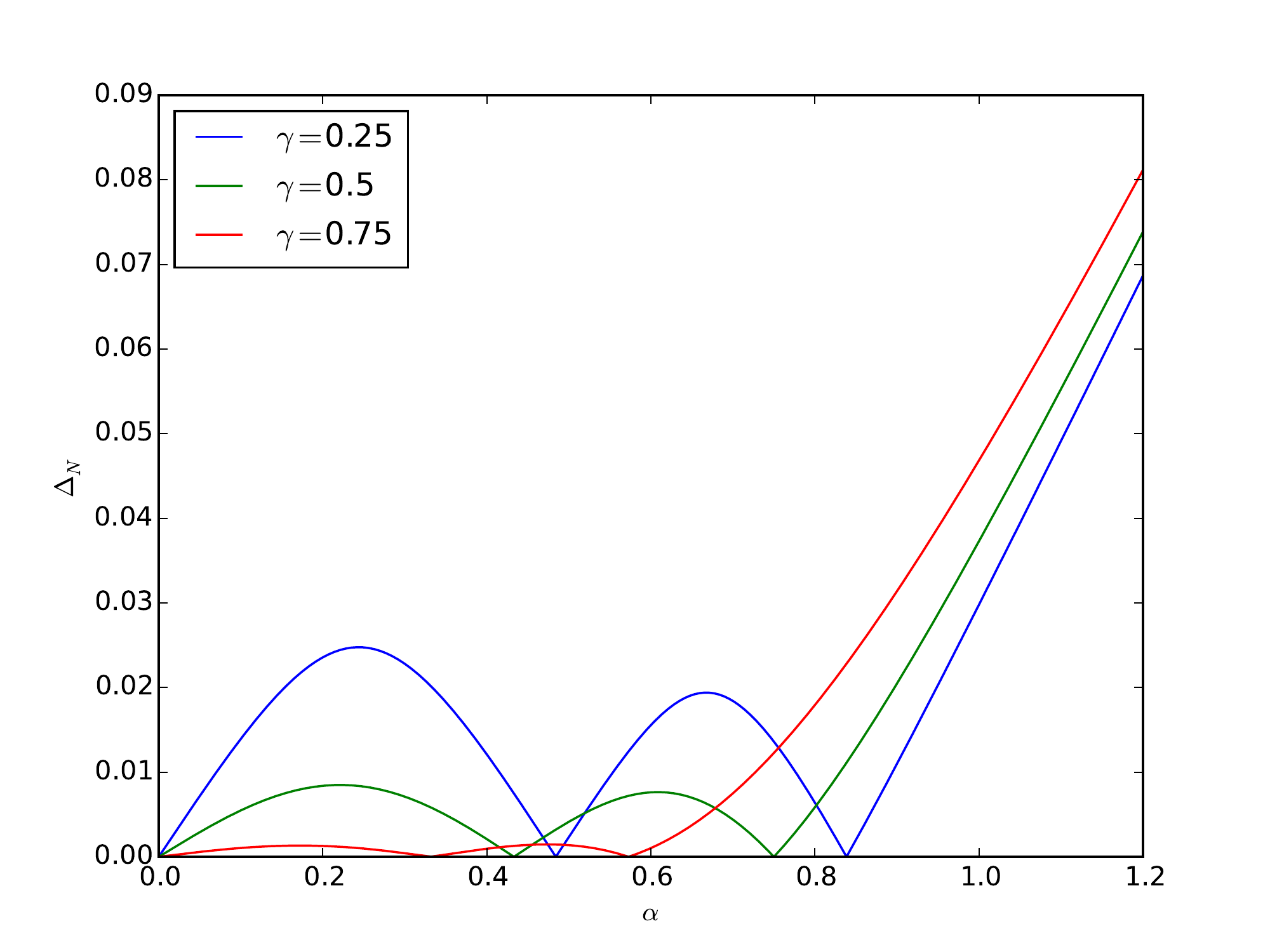}
  \caption{Energy gap for $N=5$.}
\end{subfigure}
\hspace{5mm}
\begin{subfigure}{.47\textwidth}
  \centering
  \includegraphics[width=\textwidth]{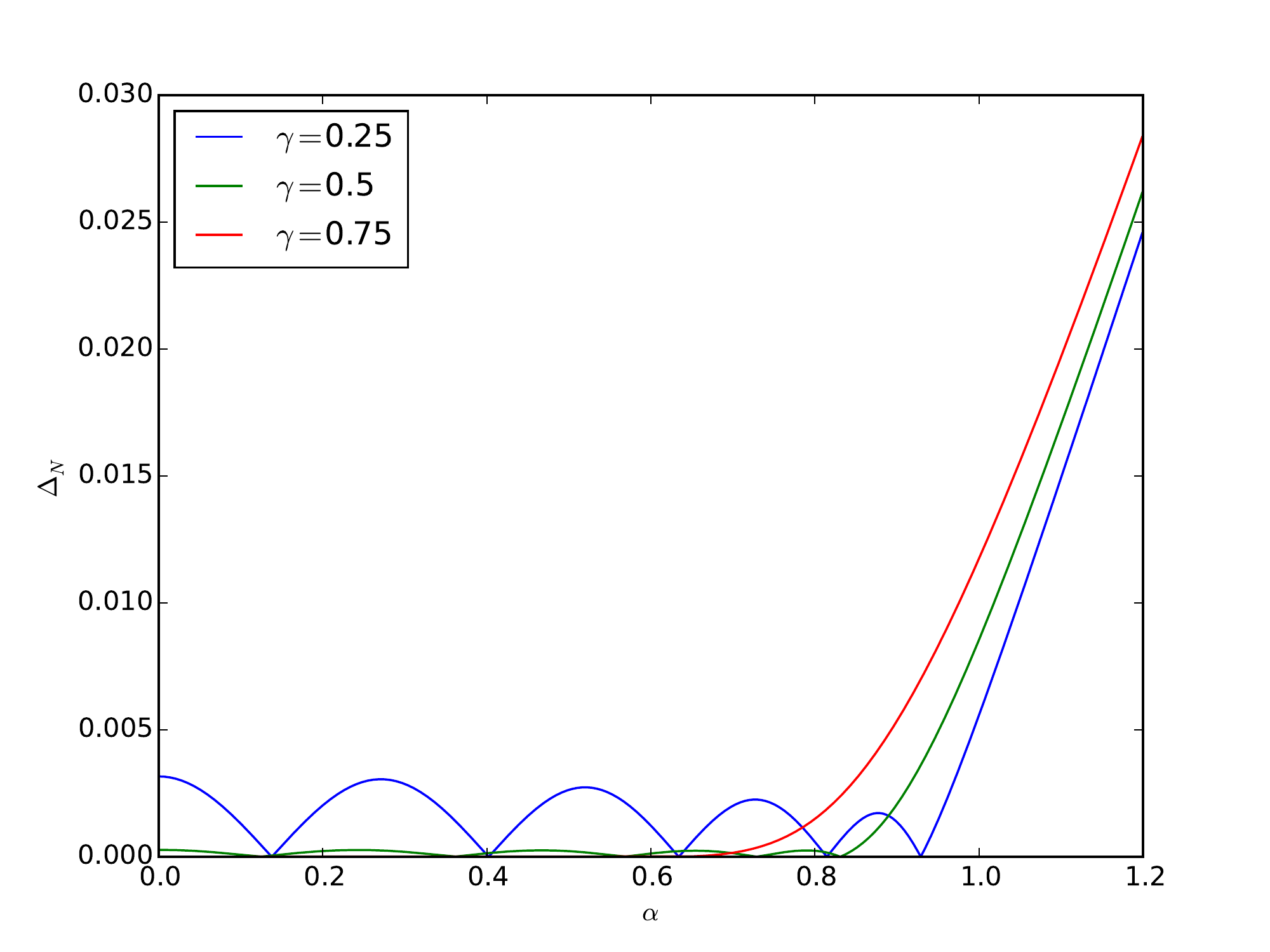}
  \caption{Energy gap for $N=10$.}
\end{subfigure}
\begin{subfigure}{.47\textwidth}
  \centering
  \includegraphics[width=\textwidth]{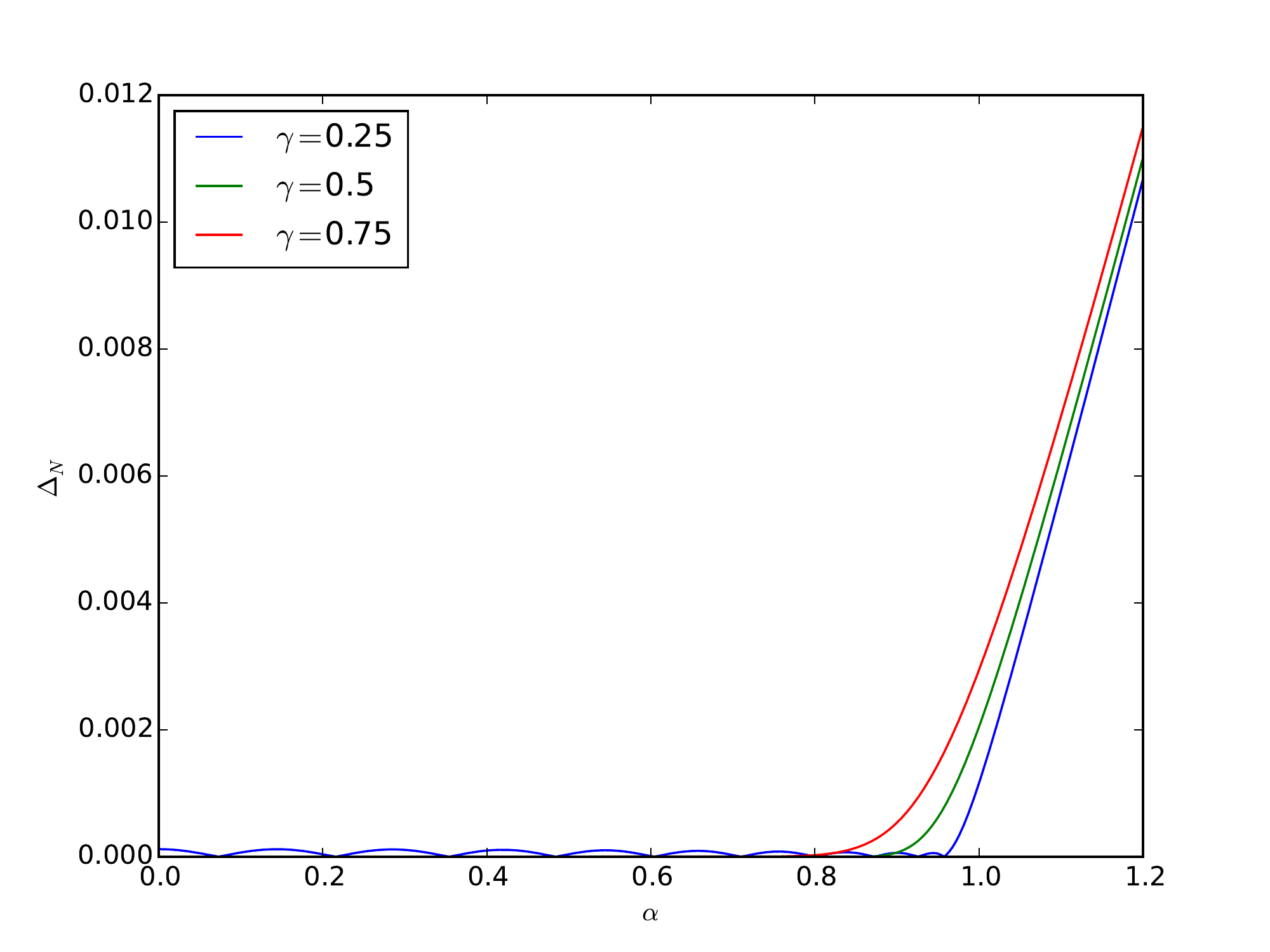}
  \caption{Energy gap for $N=20$.}
\end{subfigure}
\hspace{5mm}
\begin{subfigure}{.47\textwidth}
  \centering
  \includegraphics[width=\textwidth]{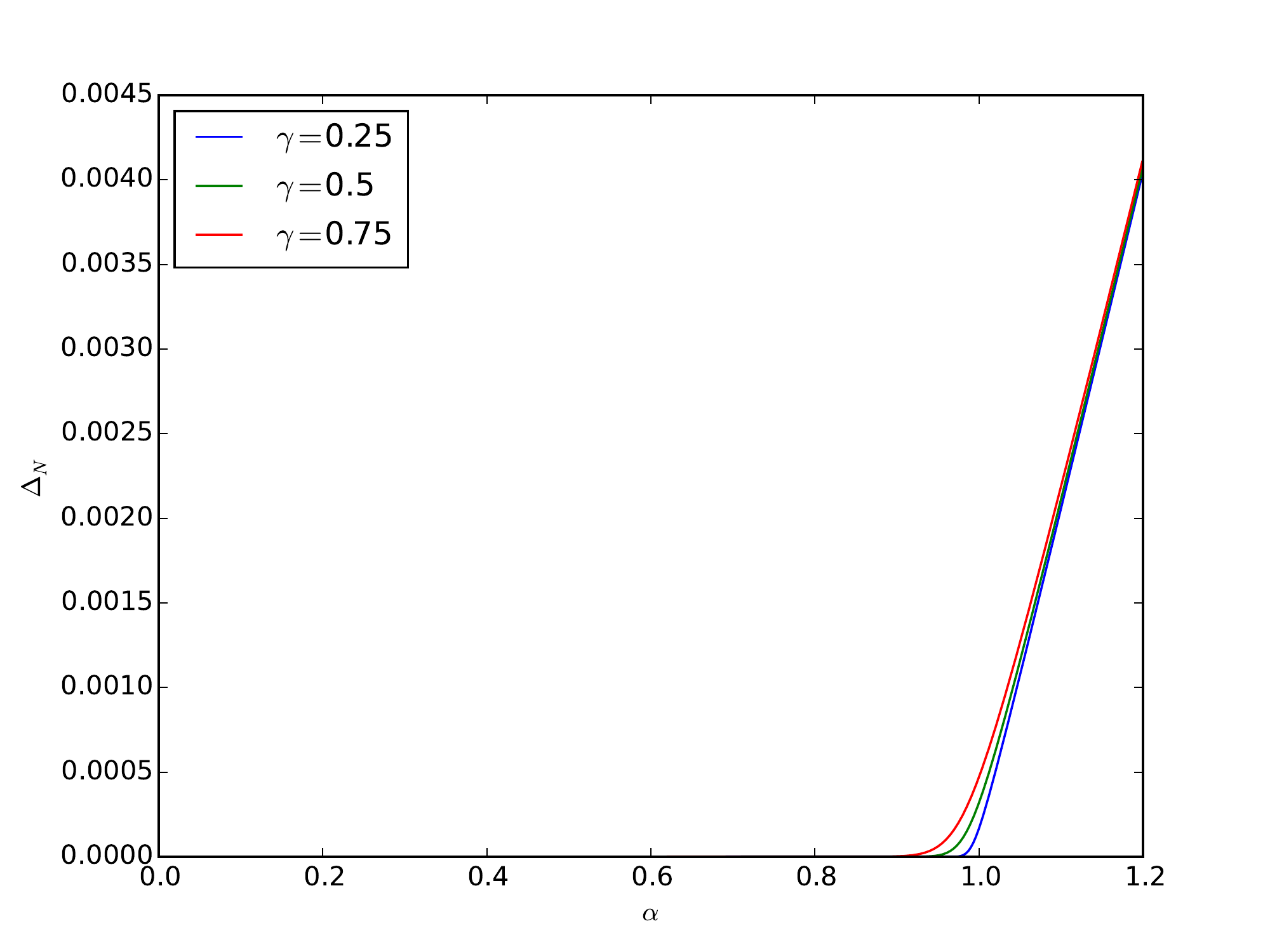}
  \caption{Energy gap for $N=50$.}
\end{subfigure}
\caption{The energy gap of an open chain for different chain lengths. The oscillatory behaviour for $\alpha<\sqrt{1-\gamma^2}$ can be seen. Moreover, with the raise of the chain length the gap tends to zero and the oscillations are faster. For a detailed description, see \cite{HvGR}.} 
\label{fig:gap_n}
\end{figure}

We have confirmed numerically that the value of energy gap, extrapolated to $N\to\infty$,  {\em is zero}. 
This observation precludes (in a numerical manner) the occurrence of the Kosterlitz-Thouless transition in the XY chain.

The distinguishing feature of K-T transition is the exponential closing of the gap together with the smoothness of ground-state energy at the transition point. In our case, energy gap is asymptotically zero in the neighbourhood of transition point, therefore the gap is not an appropriate object for testing analyticity of the ground-state energy. Intuitively, one could conjecture that the non-analyticity combined with the smoothness of ground-state energy should be somehow related to the structure of energy levels, but we are not able to specify such a relation. More precisely, we don't know how to formulate asymptotic behaviour of energy levels in the neighbourhood of the circle $\alpha^2+\gamma^2=1$, and leave this as an open problem. We only conclude that the transition between oscillating and ferromagnetic phases is not of the Kosterlitz-Thouless type.
%%%%%%%%%%%%%%%%%%%%%%%%%%%%%%%%%%%%%%%%%%%%%%%%%%%%%%%%
\section{Summary, perspectives for future work}
%%%%%%%%%%%%%%%%%%%%%%%%%%%%%%%%%%%%%%%%%%%%%%%%%%%%%%%%
We have analysed the  the formula for ground-state energy  of the  anisotropic XY model  in transverse magnetic
field. 
We have examined differentiability properties of the expression.
It turned out that the second derivative of energy (i.e. the transverse susceptibility)
is logarithmically divergent in the neighbourhood of the lines
 $\alpha=1$ and $\gamma=0$. This behaviour is characteristic for Ising-type phase transition. This fact
 has been known earlier (see for instance 
\cite{BunderMcKenzie}), but here one can obtain sub-leading exponents and amplitudes
up to arbitrary order. We have presented some of first terms of such an expansion.

An interesting open problem (as far as we know) is the question of {\em universality} of
Ising-type critical behaviour near these lines. The universality principle has been formulated
more than 40 years ago, but rigorous proofs are very rare. One of such results is
the proof of universality for 2d Ising model due to Spencer \cite{Spencer}. It would be
very interesting to adapt the technique used in \cite{Spencer} to quantum XY chain, for lines $\alpha=1$ and $\gamma=0$. On the physical grounds,
such universality is expected, but -- as far as we know -- the rigorous proof is lacking.

We have also established that the energy is differentiable up to arbitrary order
on the line
$\alpha^2+\gamma^2=1$. Again, it is interesting to check whether this behaviour is stable
 (universal) against perturbations
of Hamiltonian. Another  aspect of the smoothness of energy on the line $\alpha^2+\gamma^2=1$,
 it is interesting to confront this result with paper \cite{entropiaR},
where differentiability property of {\em Renyi entropy} has been analysed. The result there
is that derivative of Renyi entropy is discontinuous on the line $\alpha^2+\gamma^2=1$. Apparently, differentiability
properties of energy and entropy are different. It  seems strange and it would be interesting
to take a closer look at this question.

{\bf  Acknowledgments.} We are grateful to Prof. J. Stolze for drawing our attention to paper \cite{AfterStolze}. Tomasz Maci\k{a}\.{z}ek is supported by Polish Ministry of Science and Higher Education ``Diamentowy Grant'' no. DI2013 016543 and ERC grant QOLAPS.
%%%%%%%%%%%%%%%%%   BIBLIOGRAPHY     %%%%%%%%%%%%%%%%%%%%%%%%%%%%%%%%%%%%%%%

\end{document}